\documentclass[12pt]{article}
\title{FROM CONSTITUENT QUARKS TO
HADRONS\\ IN COURSE OF NUCLEAR MATTER EXPANSION \footnote{ This
work is supported in part by the Russian Foundation for Basic
Researches, grant 00-02-17250 and by Scientific School grant
00-15-96696}} \author{O.D.  CHERNAVSKAYA \enspace
\mbox{and} \enspace I.I.  ROYZEN}
\topmargin=-1cm \date{} 
\begin{document} \maketitle
\centerline{\em{Lebedev Physical Institute of RAN, Leninski
prospect 53, 117333 Moscow,Russia}} \centerline{e-mail:
$<chernav@lpi.ru>,\quad <royzen@lpi.ru>$}
\addtolength{\baselineskip}{6pt} \begin{abstract} The up-dated
three-phase concept \cite{ChF} of nuclear matter evolution in
course of cooling down - from the phase of quark-qluon plasma
(QGP) through the intermediate phase allowing for massive
constituent quarks $Q$ (valons), pions and kaons (Q$\pi$K) to the
phase of hadronic matter (H) - is exploited for the treatment of
relative hadronic yields in the central region of heavy ion
collisions. The most attention is paid to the description of the
Q$\pi$K-phase which is argued to be a gaseous one and lasts until
the valonic spacing approaches the confinement radius (at the
temperature $T_H\,\simeq$ (110 $\pm$ 5) MeV), when the valons
start fusing to be locked, in the end, within the hadrons. The
hadronic yields emerged from thermal treatment of Q$\pi$K-phase
and simple combinatorial approach to the hadronization process are
shown to fit the available experimental data from AGS, SPS and
RHIC quite well.  This provides an alternative insight into the
real origin of the observed relative hadronic yields which is (to
a considerable extent) free of the well known puzzle inherent in
some conventional models where the early chemical freeze-out is
assumed:  namely, why the gaseous thermal approach to actually
tightly packed (even overlapping) hadrons seems workable?  Many
predictions for the other hadronic yields which could be observed
at these machines as well as at LHC are given.
\end{abstract}

\newpage
{\bf Introduction}

The conventional way of treatment of hadron production in the central
rapidity region of heavy ion collisions at high energies incorporates a QGP as
the short initial phase of nuclear matter (just after the nuclei encounter each
other) which then undergoes the chiral breaking transition into hot and dense
hadronic matter, the latter evolving somehow into experimentally observed free
hadrons.  The relative yields of different species of these hadrons depend on
the equation of state (EoS) of the nuclear matter they are originated from. It
has been noticed \cite{St} that the SPS data on the hadronic yields are
quite compatible with the assumption about the early chemical freeze-out in {\it
an ideal gas of crucially modified in-medium hadrons} (their effective radii
has being assumed to be as small as $\simeq 0,3$ fm only!) at the temperature
$T_{ch}\,\simeq$ 170
MeV \footnote{The straightforward estimate shows that no nearly ideal gas of
{\it slightly} modified hadrons could exist at the relevant particle (hadron)
densities, since the normal hadron wave functions would overlap substantially.}
(which is rather close to the chiral breaking temperature itself,
$T_{c}\simeq$ 200 MeV).  The theoretical observation has been also made
\cite{C-R} that thermal description of such a type could be spread to
elucidate the AGS and SIS data as well under the assumption that chemical
freeze-out temperature $T_{ch}$ was varied properly along with baryonic chemical
potential of fireball (expanding nuclear medium), being about 110 MeV and 60
MeV for AGS and SIS, respectively. As a result, a suggestion about possible
existence of low temperature QGP has been put forward and then discussed in
some more detail \cite{C-R,M-R}.  Both attributes of this approach -
crucially modified in-medium hadrons (but still survived!) and low temperature
chiral transition -
seem rather controversial and mysterious.  That is why a feeling of
dissatisfaction remains and looking for some more consistent approach seems
quite reasonable.

In this connection, the attention seems to be called reasonably
to the rather old-fashioned notion of valons $Q$ linked properly to the
corresponding current quarks - first of all, to the valons $Q(q)$,
$q\,\equiv\,u,d$ ($m_{Q(u,d)}\,\equiv\,m_{Q(q)}\,\simeq$ 330 MeV), and $Q(s)$
($m_{Q(s)}\,\simeq$ 480 MeV) - which was widely and fruitfully
exploited before QCD was developed as a consistent field theory that had no
valons as its inalienable entities.  However, till now, QCD remains actually a
quite workable theory of hard processes, but it suffers from the lack of even
qualitative results for the processes at low and intermediate energies, i.e.,
just for those ones which proceed within nuclear matter below the chiral
breaking temperature $T_{c}$.  Thus, maybe, embedding the valons unambiguously
into the body of QCD is still waiting for its turn, although the known
endeavors met no success \cite{Wil}.  Qualitative physical motivation in
favor of possibly essential role of valonic mass scale and of an intermediate
phase, allowing for valons and separating QGP and hadronic ones, can be found
in ref.s \cite{Sh,F,...}.  It is worthy to mention that experimentally observed
(in AA collisions) significant excess in low-mass dilepton ($e^+ e^-$) yield (as
compared to pA ones) could be described quite well \cite{00} in the
framework of an approach allowing for Q$\pi$K-phase.

Recently the general idea has been put forward \cite{01} which could
provide an insight into the physical reasons underlying the unified thermal
description of yields of different hadron species observed in heavy ion
collisions.  In the present paper, we show in more detail how could valons
help in treatment of relative hadronic rates, getting rid of the above
curiosity inherent in the conventional approaches.  The theory is confronted to
the available results of AGS, SPS and RHIC, and many other ratios of hadronic
yields are predicted (for LHC as well).

{\bf General Description of the Approach}

Most probably, just after chiral symmetry breaking, the nuclear matter
is subjected for dual description \cite{00}: it can be treated either as highly
compressed "hadronic liquid" or as a state, in which the most of hadron species
can not survive and the dominant degrees of
freedom are not still hadrons but valons instead \footnote{Presence of massive
valons is just indicative for chiral symmetry breaking.}. Of course, one can choose
either. However, for taking the first way, one is supposed to know, at least,
the corresponding EoS. Who is aware of it? The basic point of the suggested
approach is that the valons which would be
produced in course of chiral symmetry breaking are to be treated reasonably
as a gas from the very beginning ({\it in contrast to the hadrons!}) due
to their {it really} small size \cite{Anis}, $r\,\simeq$ 0,3 fm.  Indeed, even
at $T$ = 170 MeV the particle density within the "ideal" Boltzman valonic
gas would be about $$ \frac{12\,T^3
}{\pi^2}\,[2(\frac{m_{Q(q)}}{T})^2K_2(\frac{m_{Q(q)}}{T})cosh(\mu_{Q(q)}/T)\,
+\,(\frac{m_{Q(s)}}{T})^2K_2(\frac{m_{Q(s)}}{T})cosh(\mu_{Q(s)}/T)]\,\simeq\,1\,\mbox{fm}^{-3}$$
where $\mu_{Q(q)}$ and $\mu_{Q(s)}$ are typical for heavy ion collisions
$Q(q)$- and $Q(s)$-valon chemical potentials,
respectively\footnote{We do not distinguish here between chemical potentials of
$Q(u)$- and $Q(d)$-valons caused by difference in proton and neutron content of
heavy colliding nuclei, although at low energies and large values of $\mu_q/T$
(SIS, AGS) it may result in a noticeable effect.}, and thus the "valonic bodies"
themselves occupy about 10\% of the total volume only. This rather
optimistic estimate noticeably worsens after taking into account some
balancing fraction of the "large-sized" pions and kaons which are
produced inevitably as the chemical equilibrium sets in \footnote{All other
hadronic species possess too large masses and lose their identity within this
color conducting medium, being very unstable with respect to decay into valons,
especially due to collisions; see, e.g., ref.  \cite{Ele} for a more consistent
treatment of $\rho$-meson degradation along with increase of nuclear matter
density.}, see Fig. 1.  Fortunately, this hadronic fraction never exceeds 25\%
(see below), and thus this three-component Q$\pi$K state is still to be treated
reasonably as a gas \footnote{This is especially true near the hadronization
temperature $T_H$, what is most relevant in the context of the problem under
discussion.}, although quasi-ideality of this gas becomes somewhat more
questionable.  That is why below we proceed as far as possible with a more
general consideration based on the chemical kinetics only, and then confront
the results to what one can obtain in the selfconsistent ideal gas
approximation. We will see {\it a'posteriory} that the results emerged from
these two ways of doing are quite compatible.

Now, the general pattern of evolution of the hot
nuclear fireball produced in course of heavy ion collisions looks as follows.
While expanding and cooling down from $T_{c}$ to $T_H$,
the nuclear matter is gradually enriched with pions and kaons and impoverished
with valons.  Situation changes
dramatically when mean spacing of valons scales up to become about the
confinement radius.  As a result, gradually increasing color screening length
approaches the critical value and the dominant degrees of freedom become no
longer valons (since they are getting not free) but the hadrons themselves,
i.e., the bulk of hadrons is to build up.  This is a phase transition, if
the hadronization proceeds near the same temperature $T\,\simeq\,T_H$;
otherwise, this is a phase crossover.  The lattice calculations suggest that
peculiarities of phase transition and even its occurrence itself is regulated by
the value of chemical potential (or net baryonic density) within nuclear
matter. We do not touch here this delicate question and concentrate on the
pattern linked to a certain phase transition. This pattern is shown below
to face no obvious contradictions with the available observables
within the nowday accuracy of the data (see the Table). The relevant diagrams of hadron
production are depicted in Fig.  2.

{\bf Calculations and Results}

At the hadronization phase transition ($T\,=\,T_H$), all hadron species
are produced, generally, in the same manner.
Unlike the Q$\pi$K-phase, the correlation length remains large comparing to the
mean particle spacing (even "infinite", if it is the second order one)
under the whole this phase transition long.
Therefore, multiparticle interactions dominate at
this stage. One can assume reasonably also that now, once being created, all
the hadrons survive because the confinement starts working.

Thus, the final
(observable) pions (kaons) are produced in the two-fold way: first, there are
{\it equilibrium} pions (kaons) mentioned above (those, which are "stored up"
in course of Q$\pi$K-phase evolution, before overall hadronization, see Fig. 1),
and second, there are pions coming from $Q(q)\bar Q(q)\,\to\,\pi\,+\,X$
reactions (kaons coming similarly from $Q(q)\bar Q(s)$ and $Q(s)\bar Q(q)$
reactions) just at $T\,=\,T_H$.  The latter are produced at the stage of
hadronization only, just like the other hadron species, see Fig. 2.  The number
of the former pions reads obviously \begin{equation}
n_{\pi}\,\simeq\,\frac{bn}{1\,-\,b} \end{equation} where $n$ is the total
number of color particles within fireball and $b$ is the pionic fraction which
is calculated in the Appendix (see eq.s A(1)-A(6)). The numerical value of $b$ turns
out to be rather stable for SPS, RHIC and LHC:  $b\,\simeq$ 0,22, whereas for
AGS it is considerably lower: $b\,\simeq$ 0,13. Making use of the diagrams
shown in Fig.  1(b), 1(c), the number of former kaons is estimated quite similarly
(see eq.  A(7)) to be \begin{equation} n_{K^+}\,\simeq\,n_{K^0}\,\simeq
\frac{4n_{\pi}\,+\,3n_{Q(q)}}{6(4n_{\pi}\,+\,3n_{\bar Q(q)})}\,n_{\bar
Q(s)} \end{equation} where $n_{\bar Q(s)}$ is the number of
($\bar Q(s)$ valons.  The number of $K^-$ ($\bar K^0$) is obtained by
replacement here valons $\leftrightarrow$ antivalons.

As for the latter pions, one can express their number by tracing a valon $Q(q)$
(antivalon $\bar Q(q)$) within hadronizing nuclear medium. While moving along
the mean free path, it coalesces with an antivalon (valon) with the
probability \footnote{Of course, at the hadronization stage, any valonic
interaction which could result in producing a hadron is reasonably assumed to do
produce it.  Besides, the valon-hadronic interaction is insignificant at this
stage.} $n_{\bar Q(q)}/n$ \enspace ($n_{Q(q)}/n$), the total rate of such collisions
thus being $n_{Q(q)} n_{\bar Q(q)}/n$. In each collision, $\pi$-meson plus $X$
are produced, "$X$" allowing possibly for a number $j$ of pions too, see Fig. 2
(the latter is also true for the processes of production of other hadrons).
Thus, accounting eq.  (1), the total negative pion yield (which is
assumed to be equal to 1/3 of the total rate of pions) reads \begin{equation}
N_{\pi^-}\,\simeq\, \frac{1}{3}\,[\frac{bn}{1\,-\,b}\,+\,\frac{
(1\,+\,\langle j\rangle)n_{Q(q)}
n_{\bar Q(q)}}{n}]\,+\,\frac
{\langle j\rangle}{3}[N_B\,+\,N_{\bar B}\,+\,(N_K\,-\,n_K)\,+\,...]
\end{equation} where $N_B\,(N_{\bar B})$ and $N_K$ are the baryon (antibaryon)
and total $K$-meson
yields, respectively, and $\langle j\rangle$ is the mean value of $j$ which is
easily estimated to be 0 $\leq\,j\,\leq$ 1 because of the phase space
limitation \footnote{The results appear to be rather insensitive to the value of
$\langle j\rangle$ within this domain.} (points stand for a number of small
terms which are meaningless
within the accuracy of both the theory and the data).  Proceeding similarly and
making use of eq. (2), one obtains for
$K$-meson yield:  \begin{eqnarray}
N_{K^+}\,\simeq\,N_{K^0}\,\simeq\,\frac{
4n_{\pi}\,+\,3n_{Q(q)}}{6(4n_{\pi}\,+\,3n_{\bar
Q(q)})}\,n_{\bar Q(s)}\,+\,\frac{n_{Q(q)}
n_{\bar Q(s)}}{2n},
\\N_{K^0_S}\,\simeq\,N_{K^0_L}\,\simeq\,0,5(N_{K^0}\,+\,N_{\bar K^0}),
\end{eqnarray} $N_{K^-}$ and $N_{\bar K^0}$ being obtained as before (see eq.
(2)). The yields of some other mesons are easily estimated by means of the
similar combinatorial and/or relevant cross-section consideration. For example,
\begin{equation} N_{\phi}\,\simeq\,\frac{n_{Q(s)}n_{\bar Q}(s)}{n}
\end{equation} As for $\eta$-meson, the relative rate $N_{\eta}/N_{\pi^0}$ can
be immediately calculated by comparing the corresponding cross sections, since
two channels involved ($Q(q)\bar Q(q)\,\to\,\pi \pi^0$ and $Q(q)\bar
Q(q)\,\to\,\pi \eta$, see Fig. 2(b)) {\it directly} compete with each other.  In
addition to distinction in masses of these mesons which results in about
(2$\div$3)-time contraction (at the relevant energies/temperatures) of the $\pi
\eta$ final phase space (as compared to the $\pi \pi^0$ one), one should take
into account that no $\rho$-meson intermediate state is allowed in $\pi^{\pm}
\eta$ channel.  Being integrated over the energy, the corresponding excess in
the $\pi \pi$ yield results in an extra factor about 4$\div$5, thus making
finally $N_{\eta}/N_{\pi^0}\,\simeq$ (0,08$\div$0,1). Unfortunately, the
flexibility of this estimate is rather large.

Some additional peculiarities are to be accounted for getting the
formulae for the baryon rates. First, the factor
$(1\,+\,2n^2_Q(q)/n^2)^{-1}$ (or $(1\,+\,2n^2_{\bar Q(q)}/n^2)^{-1}$)
should enter the formulae to eliminate the double counting of a "propagating"
(anti)valon and a pair of the identical (anti)valons it encounters
(actually, this factor may be significant for (anti)nucleon production only);
and second, one should keep in mind that what is taken experimentally as
$\Lambda (\bar \Lambda)$ is actually $\Lambda (\bar \Lambda)\,+\,\Sigma^0 (\bar
\Sigma)$ (since $\Sigma^0\, (\bar \Sigma^0)$ decays into $\Lambda (\bar
\Lambda)\,+\,\gamma$ in $\simeq\,10^{-19}s$), and that the final
momentum phase spaces for $\Lambda$ and $\Sigma$ production differ from each
other (compare to the above remark on $\eta/\pi^0$-yield ratio) as well as the
combinatorial factors for production of the neutral and charged hyperons
(see the diagram $h$ in Fig.  2)\footnote{As a result, the rate
of $\Lambda\, (\bar \Lambda)$ production turns out to be almost twice as large
as that of both $\Sigma^{\pm}\, (\bar \Sigma^{\pm})$-meson species.}. Thus, we
arrive at \begin{eqnarray} N_p \,\simeq\,N_n\,\simeq\,\frac{
n^3_{Q(q)}}{2n^2(1\,+\,2n^2_{Q(q)}/n^2)},\qquad N_{\bar p}\,\simeq\,N_{\bar
n}\,\simeq\,\frac{n^3_{\bar Q(q)}}{2n^2(1\,+\,2n^2_{\bar Q(q)}/n^2)}\\
N_{\Lambda}\,+\,N_{\Sigma^+}\,+\,N_{\Sigma^-}\,\simeq\,
\frac{n_{Q(s)}n^2_{Q(q)}}{n^2}\,\simeq\,1,6N_{\Lambda}\\
N_{\bar \Lambda}\,+\,N_{\bar \Sigma^+}\,+\,N_{\bar
\Sigma^-}\,\simeq\,\frac{n_{\bar Q(s)} n^2_{\bar
Q(q)}}{n^2}\,\simeq\,1,6N_{\Lambda}\\
N_{\Xi^0}\,\simeq\,N_{\Xi^-}\,\simeq\,\frac{n_{Q(q)}n^2_{Q(s)}}{2n^2}, \qquad
N_{\bar \Xi^0}\,\simeq\,N_{\bar
\Xi^+}\,\simeq\,\frac{n_{\bar Q(q)}n^2_{\bar Q(s)}}{2n^2}\\
N_{\Omega^-}\,\simeq\,\frac{n^3_{Q(s)}}{n^2}, \qquad
N_{\bar \Omega^+}\,\simeq\,\frac{n^3_{\bar Q(s)}}{n^2},
\end{eqnarray} and so on.

The comparison of the above predictions for baryon and antibaryon rates with the
experimental data asks for a more careful
discussion (especially, for antibaryons at AGS).  Theoretical consideration of
antibaryon production is a subtle point of any model because it implies the
details of the evolution dynamics of hadronizing nuclear matter . In contrast
to mesons, which, once being produced in course of hadronization, preserve their
identity (undergo predominantly elastic scattering only), the (anti)baryons
start to annihilate right away after production; the lower is $T_H$, the larger
is the annihilation cross section.  A rather high
density of surrounding baryons (large chemical potential at AGS) and very large
annihilation cross section, $\sigma_{p\bar p}\,\simeq$ 100 mb, favor to very
strong absorption which crucially
diminish the antibaryon number before hadrons scatter away and stop
interacting.  Then, what remains passes through the detectors. Of course, this
effect is missed in the above formulae.  Since any theoretical estimate of this
absorption could be only qualitative and is far from being reliable, we try to
draw it directly from the experimental data.  We call the attention to the fact
\cite{E802}  that measured at AGS  $\bar p/\pi^+$ yield in low participant number
($\simeq$ 20) collisions is about twice as large as it is in the high
participant number ($\simeq$ 80$\div$100) ones.  Meanwhile, the antiproton
production is
a rare process and thus it is expected to rise up proportionally
to the number of participants squared. If so, then one would expect that
relative $\bar p/\pi^+$ yield
should increase linearly. That is why one can deduce that
annihilation eats at AGS about up to 90\% of the initially produced
antiprotons. At SPS, the relevant ratio of yields is $^{\cite{NA49}}$ about 0,5
and thus annihilation eats (if any) not more than 50\% of
antiprotons. This is due to both substantial increase of the pionic
fraction amongst the producing hadrons and decrease of baryonic chemical
potential (both "dilute" the baryonic content and make lower the probability of
baryon-antibaryon collisions).  At RHIC and LHC, absorption of (anti)baryons
seems getting, most probably, out of the game.

Below, the results are presented in the Table (see the column $th_1$) and
Fig.  3.  The ratios $\frac{n_{Q(q)}}{n_{\bar Q(q)}},\,\frac{n_{Q(s)}}{n_{\bar
Q(s)}},\,\frac{n_{Q(q)}}{n_{Q(s)}}$ and value of $\langle j\rangle$ were found
by minimization of the sum of squared deviations:
$$\chi^2\,\simeq\,\sum\limits_{i=0}^{i=k}
\,(1\,-\,\frac{a^i_{th}}{a^i_{exp}})^2$$ were $k$ is the number of different
processes under consideration.

Note, that till now no specific assumptions about the properties of Q$\pi$K
gas were made.  The assumption of its quasi-ideality becomes unavoidable, when
we try to extract the hadronization temperature $T_H$ from the ideal-gas
formulae:$$\frac{n_{Q(q)}}{n_{\bar Q(q)}}\,=\,exp(2\mu_{Q(q)}/T),\quad
\,\frac{n_{Q(s)}}{n_{\bar Q(s)}}\,=\,exp(2\mu_{Q(s)}/T)$$ and
$$\frac{n_{Q(q)}}{n_s}\,\simeq\,(\frac{m_{Q(q)}}{m_{Q(s)}})^{3/2}\,exp[
\frac{(m_{Q(s)}\,-\,m_{Q(q)})}{T}]\, exp [\frac{
(\mu_{Q(q)}\,-\,\mu_{Q(s)})}{T}]$$
This way of doing results in nearly the same temperature, $T_H\,\simeq$
(110$\pm$5) MeV for AGS, SPS and RHIC.
Predictions for LHC given in the Table are obtained for the same value of $T_H$
at the point $\mu_i\,=\,\mu\,=\,0$, see Fig.  3.

{\bf Discussion and Concluding Remarks}

It is worth mentioning that the above results for SPS are quite similar to those
which one can carelessly obtain in the framework of the ideal gas approximation
for description of Q$\pi$K-phase instead using the detailed balancing equations
(compare the columns $th_1$ and $th_2$ in the Table. This suggests that either
the valon-pion-kaon gas is really quasi-ideal or the relative content of
different components in this gas is a crude characteristic (this seems
quite plausible) which is rather insensitive to its fine tuning. Anyway,
the compatibility of the results seems indicative for the validity of the
approach itself.

The value of hadronization temperature, $T_H\,\simeq$ 110 Mev, we have found
is quite close to the value of thermal freeze-out temperature which has
been estimated \cite{NUXU} for SPS and RHIC to be $T_f\,\simeq$
(100$\div$110) MeV by fitting the transverse momentum spectra within the
framework of the hydrodynamical model. If so, then fireball evolution from
$T_H$ to $T_f$ takes a rather short time (of some few Fermi), and one can
qualitatively understand, why the antiproton absorption in course and just
after hadronization is crucially significant for AGS only (when the relative
content of the nucleons is very high). In particular, that is why we do not
need to look for a special mechanism of levelling the absorption in contrast to
what should be necessarily invented \cite{RSh} in the schemes with early
chemical freeze-out.

Still one point which deserves mentioning is the approximate coincidence of the
hadronization temperatures for AGS, SPS and RHIC, although the relevant
chemical potentials differ substantially and thus the relevant EoS's are
expected to be different as well. Actually, we can not insist on the validity
of this result: $\chi^2$ minimization is not the meaningful test of
the theory for AGS and RHIC, since the number of parameters to be varied is
only slightly less, than the number of the necessary data from these
accelerators available at present. In addition, one should not take
seriously too low values of $\chi^2$ because of a rather insufficient accuracy
of the data themselves.

Possible variation of hadronization temperature along with valonic chemical
potential (and thus with colliding ion energy) is closely related to the very
interesting hypothesis \cite{F} that color, namely valonic, deconfinement is
reasonably expected to take place already at very low interaction energies
(ion kinetic energy being about 300 MeV only!) because the nucleonic density is
only about twice as large as the nucleus one, and one should make a rather
weak effort for getting valons unable to "recognize their original nucleons".
If it is the case, then, in particular, a thermal treatment of hadron
production should be applicable even at such a low energies. This would provide
the direct indication on validity of the very notion of valon itself . Of
course, a high-luminosity machine equipped with high-precision detectors is
needed to try looking for the relevant manifestations.

Thus, our conclusion is that the suggested approach provides quite successive
treatment of the available data for the hadron yields
as well as for the low-mass dilepton production\cite{00} in heavy ion collisions.
At the same time, it is
free of the crucial inconsistency inherent in some conventional approaches,
although some significant questions still remain which are unanswered.

This is our pleasure to express our deep gratitude to E.L. Feinberg for the
invaluable discussions and permanent interest to the work.

\newpage

APPENDIX
\renewcommand{\theequation}{A.\arabic{equation}}
\setcounter{equation}{0}

1. Being averaged over the particle distributions, the detailed
balancing equation reads:  \begin{equation}\label{A1} \nu_{Q(q)}(T)\,\overline
{\Omega_{\pi}}(T)\,\simeq\,\nu_{\pi}(T)\,\overline{\Omega_{Q(q)}}(T)
\end{equation}\label{A1} where $\overline{\Omega_i}$ are the mean values of the
corresponding final state phase spaces\footnote{The binary reactions are to be
considered only because 2 $\to$ 4 reactions are substantially suppressed by
scarcity of the typical thermal final state phase space at $T\,\simeq\,T_H$.}.
Since each antivalon ${\bar Q(q)}$ of a certain color and flavor encounters a
valon $Q(q)$ with the probability $n_{Q(q)}/(n\,+\,n_{\pi})$, the rate of
$Q(q)\bar Q(q)$ collisions is \begin{equation}\label{A2} d\nu_{Q(q)}
\,=\,\frac{(1\,-\,b)n_{Q(q)} n_{\bar Q(q)}}{n}\,\frac{dt}{\langle t\rangle},
\end{equation}\label{A2}
{$\langle t\rangle$} being the mean free time between the successive collisions.
Quite similarly, a $\pi^0$-meson encounters
another $\pi$-meson with the probability
$b$, the total rate of $\pi^0 \pi$ collisions being, therefore,
$\frac{2}{9}\,bn_{\pi}\,\frac{dt}{\langle t\rangle}$ ($\pi^0 \pi^{\pm}$
collisions) plus $\frac{bn_{\pi}}{18}\,\frac{dt}{\langle t\rangle}$ ($\pi^0
\pi^0$ collisions); the rate of $\pi^+ \pi^-$ collisions is obviously
$\frac{bn_{\pi}}{9}\,\frac{dt}{\langle t\rangle}$. Of course, $\pi^+\pi^+$
and $\pi^-\pi^-$ collisions are out of the game in the detailed balancing
principle equation (within the above approximation), since they never result in
a two-valonic final state.  Thus, for the total rate of $\pi \pi$ collisions to
be accounted one gets \begin{equation}\label{A3}
d\nu_{\pi}\,=\,\frac{7}{18}\,bn_{\pi}\,\frac{dt}{\langle t\rangle}
\end{equation}\label{A3}
The averaged valonic and pionic phase spaces are \begin{equation}\label{A4}
\overline{\Omega_{Q(q)}}\,\simeq\,N^2_f (2S_Q +\,1)^2
N_c\,\overline{p^2_{(Q(q)}} \quad \mbox{and} \quad \overline
{\Omega_{\pi}}\,\simeq\,(2I_{\pi}+\,1)^2\overline{p^2_{\pi}}
\end{equation}\label{A4} respectively, where $S_Q$ is the valonic spin and
$I_{\pi}$ is the pionic isospin, $p_i$ is the valonic or pionic momentum in
the CMS of two interacting valons or pions, $N_c$ and $N_f$ are the color and
flavor numbers ($N_c$ stands in eq. (A3) instead of $N_c^2$, since only the
color-singlet sector of the total two valon phase space is allowed for).  The
straightforward averaging over the Boltzmann distribution gives for the mean
energy of a particle with mass $m$:
\begin{equation}\label{A5}\overline{E^2 (m,T)}\,=
\,T^2\,[\,3\frac{\frac{m}{T}\,K_1(\frac{m}{T})}{K_2(\frac{m}{T})}\,+\,12\,+\,
\frac{m^2}{T^2}\,]\end{equation}\label{A5}
where $K_{1,2}$ are the corresponding Bessel functions. The CMS value of
$\overline{p^2_{\pi}}$ ($\overline{p^2_{Q(q)}}$) of each particle in the pionic
(valonic) final state is obtained obviously by insertion into this expression
$m\,=\,m_{Q(q)}$ ($m\,=\,m_{\pi})$, subtraction $m^2_{\pi}$ ($m^2_{Q(q)}$) and
taking a half of this difference.  Combining eq.s (A1)-(A5), we easily obtain
\begin{equation}\label{A6}
b\,\simeq\,[\,1\,+\,2,5\,
(\frac{n_q n_{\bar
q}}{n^2}\,\frac{\overline{p^2_{\pi}}}{\overline{p^2_{Q(q)}}})^{-1/2}]^{-1}
\end{equation}\label{A6} where the values of $\overline{p^2_i}$ are to be
taken in the accordance with eq. (A5).

Now, consider the equilibrium kaon fraction
and derive eq. (2)
for the in-Q$\pi$K-medium $K^{0,+}$-meson  yields. For doing that,
one should take
into account the diagram shown in Fig. 1(b), 1(c). It is easy to check that,
at the
relevant temperatures, the averaged CMS momenta squared are nearly equal for
all four states under consideration, and thus the corresponding detailed
balancing equation reads\footnote{Of course, the matrix element moduli for two
processes involved are assumed to be nearly equal.}:

\begin{equation}\label{A7}
\frac{(2I_K\,+\,1)}{(2I_{\pi}\,+\,1)}\,n_{\pi}n_{\bar
Q(s)}\,+\,\frac{(2I_K\,+\,1)(2I_{\pi}\,+\,1)}{(2S_Q\,+\,1)^2}\frac{n_{\bar
Q(s)}n_{Q(q)}}{N_c
}\,\simeq\,\frac{(2I_{\pi}\,+\,1)}{(2I_K\,+\,1)}\,n_{\bar Q(q)}n_K
\end{equation}
$$+\frac{(2S_Q\,+\,1)^2\,N_c}{(2I_K\,+\,1)(2I_{\pi}\,+\,1)}\,n_{\pi}n_K$$ where
$K$ stands for $K^+$ or $K^0$ and $I_i$ is the relevant isospin. This gives eq.
(2).

Making use of these formulae, one can easily estimate that the $K$-meson
induced correction to the value of $b$ is really quite small: say, for SPS,
$\frac{\delta b}{b}\,\simeq\,+(4\div5)\,10^{-2}$ and it is undoubtedly absorbed
into inaccuracy of the approach itself.

{\bf Figure captions}

Fiq. 1. The processes taken into account for getting the detailed
balancing equations (see Appendix) which regulates the relative
content of different components in Q$\pi$K gas. Diagrams (b) and (c) refer,
actually, to the same diagram linked to the two cross channels.

Fig. 2. The typical processes which proceed at the hadronization phase
transition ($T\,\simeq\,T_H$). The dotted and wave lines refer to a number of
the attendant pions and gluons, the latter being absorbed by medium.

Fig. 3. The ($\mu T$) phase diagram allowing for the Q$\pi$K phase. The solid
line (1) refers to the QGP $\to$ Q$\pi$K phase transition, the strip (2)
between two dashed lines marks predicted in the present work temperature
interval around $T$ = 110 MeV for the expected value of Q$\pi$K $\to$ H
phase transition temperature $T_H$, and two lines (3)  are quoted  from the ref.
\cite{C-R} where they refer to the strip found by the authors for the suggested
unified description of the {\it hadronic} chemical freeze-out after the
{\it direct} QGP $\to$ H phase transition.


\begin{thebibliography}{9}
\bibitem{ChF} O.D. Chernavskaya, E.L. Feinberg, in Hot Hadronic Matter: Theory
and Experiment, Ed. J.letessier, J.Rafelski,  Plenum Press (1995);
O.D. Chernavskaya, E.L. Feinberg, J.Moscow Phys.Soc. (1996), vol.6. p.37;
E.L. Feinberg, Proc. of 2-d Int. Sakharov Conf.
Singapore: World Scientific, Ed. I.M. Dremin, A.M. Semikhatov, 1997, p.6.
\bibitem{St} P.  Braun-Munzinger, I. Heppe, J. Stachel, nucl-th/9903010 v2.
\bibitem{C-R} J.
Cleymans, K.  Redlich, Nucl. Phys. 1999 vol. A661, p. 379; nucl-th/9903063.
\bibitem{M-R} B.  Muller, J. Rafelski, Physics World 1999, vol. 12, no. 3, p.
23.
\bibitem{Wil} K.G. Wilson, Phys. Rev.  1974, vol. D10, p. 2245; K.G.
Wilson and D.G.  Robertson, hep-th/9411007.  \bibitem{Sh} E.V.  Shuryak, Phys.
Lett.  1981, vol.  B107, p.  103.  \bibitem{F} E.L. Feinberg, Sov.  Phys.
Uspecki, 1983, vol. 26, p. 1, and ref,s therein; "On Deconfinement of
Constituent and Current Quarks in Nucleus-Nucleus Collisions", Preprint of
Lebedev Physical Institute No. 197 (1989) ; in:  "Relativistic Heavy Ion
Collisions", Ed.  L.P.  Chernai, D.D.  Strottman, World Scientific, 1991.
\bibitem{...} A.V.  Bochkarev, M.E.  Shaposhnikov, Nucl.  Phys.  1986, vol.
B268, p. 220; J.  Cleymans, K.  Redlich, H.  Satz, E.  Suhonen, Z.Phys.  1986,
vol. C33, p. 151; D.V.  Anchishkin, K.A.  Bugaev, M.I.  Gorenstein, Z.Phys.
1990, vol. C45, p.  687; S.P. Baranov, L.V.  Fil'kov, Z.Phys. 1993, vol. C57,
p.149.  O.D.  Chernavskaya, E.L. Feinberg, in:  "Hot Hadronic Matter:  Theory
and Experiment", Ed. J.  Letessier, J. Rafelski, Plenum Press, 1995; J.  Moscow
Phys. Soc. 1996, vol. 6, p. 37.  \bibitem{00} O.D.  Chernavskaya, E.L.
Feinberg, I.I. Royzen,  Sov.  Nucl. Phys. 2002, vol. 65, p. 1, hep-ph/0011009.
\bibitem{01} O.D. Chernavskaya, E.L.  Feinberg, I.I. Royzen, hep-ph/0101063.
\bibitem{Anis} V.V.  Anisovich et al., Sov.  Phys.  Uspecki 1984, vol.  4, p.
553;  B.L. Ioffe, V.A.  Khoze: "Hard Processes", vol. 1:
"Phenomenology, Quark-Parton Model", North Holland, Amsterdam, 1984.
\bibitem{Ele} V.L.  Eletsky et al., Eur.  J. Phys.  1998, vol.  A3, p. 381;
Nucl. Phys. 1999, vol.  A661, p. 514c.
  \bibitem{E802} Y. Akiba for the E802 Collaboration, Nucl.
Phys. 1996, vol.  A610, p. 139c.  \bibitem{NA49} F.  Sikler for the NA49
Collaboration, Nucl.  Phys. 1999, vol.  A661, p. 45c.
\bibitem{NUXU} Nu Xu, Talk at the Conf. "Quark Matter 2001", Stony Brook,
Jan. 15-20, 2001.
\bibitem{RSh} R. Rapp and E.V. Shuryak, hep-ph/0008326 v3.
\end{thebibliography}
\end{document}